# Characterization of SABRE crystal NaI-33 with direct underground counting

M. Antonello[1], I. J. Arnquist[2], E. Barberio[3], T. Baroncelli[3], J. Benziger[4], L. J. Bignell[5], I. Bolognino[1,6],
F. Calaprice[7], S. Copello[8,9,15], I. Dafinei[10], D. D'Angelo[1,6], G. D'Imperio[10,a], M. D'Incecco[9], G. Di Carlo[9],
M. Diemoz[10], A. Di Giacinto[9], A. Di Ludovico[7], W. Dix[3], A. R. Duffy[11,12], E. Hoppe[2], A. Ianni[9], M. Iannone[10],
L. Ioannucci[9], S. Krishnan[12], G. J. Lane[5], I. Mahmood[3], A. Mariani[8], S. Milana[10], J. Mould[11,12], F. Nuti[3],
D. Orlandi[9], V. Pettinacci[10], L. Pietrofaccia[7], S. Rahatlou[10,13], F. Scutti[3], M. Souza[7], A. E. Stuchbery[5], B. Suerfu[7,16],
C. Tomei[10], P. Urquijo[3], C. Vignoli[9], A. Wallner[5], M. Wada[7,17], A. G. Williams[14], A. Zani[1], M. Zurowski[3]

[1] INFN-Sezione di Milano, 20133 Milan, Italy
[2] Pacific Northwest National Laboratory, Richland, WA 99352, USA
[3] School of Physics, The University of Melbourne, Melbourne, VIC 3010, Australia
[4] Chemical Engineering Department, Princeton University, Princeton, NJ 08544, USA
[5] Department of Nuclear Physics, The Australian National University, Canberra, ACT 2601, Australia
[6] Dipartimento di Fisica, Università degli Studi di Milano, 20133 Milan, Italy
[7] Physics Department, Princeton University, Princeton, NJ 08544, USA
[8] Gran Sasso Science Institute, 67100 L'Aquila, Italy
[9] INFN-Laboratori Nazionali del Gran Sasso, 67100 Assergi (L'Aquila), Italy
[10] INFN-Sezione di Roma, 00185 Rome, Italy
[11] ARC Centre of Excellence for All-Sky Astrophysics (CAASTRO), Sydney, Australia
[12] Centre for Astrophysics and Supercomputing, Swinburne University of Technology, Hawthorn, VIC 3122, Australia
[13] Dipartimento di Fisica, Sapienza Università di Roma, 00185 Rome, Italy
[14] The University of Adelaide, Adelaide, SA 5005, Australia
[15] Present Address: Dipartimento di Fisica, Università degli Studi di Genova and INFN Genova, 16146 Genoa, Italy
[16] Present Address: Department of Physics, University of California, Berkeley, Berkeley, CA 94720, USA
[17] Present Address: Nicolaus Copernicus Astronomical Centre of the Polish Academy of Sciences, 00-716 Warsaw, Poland

**Abstract** Ultra-pure NaI(Tl) crystals are the key element for a model-independent verification of the long standing DAMA result and a powerful means to search for the annual modulation signature of dark matter interactions. The SABRE collaboration has been developing cutting-edge techniques for the reduction of intrinsic backgrounds over several years. In this paper we report the first characterization of a 3.4 kg crystal, named NaI-33, performed in an underground passive shielding setup at LNGS. NaI-33 has a record low $^{39}$K contamination of $4.3 \pm 0.2$ ppb as determined by mass spectrometry. We measured a light yield of $11.1 \pm 0.2$ photoelectrons/keV and an energy resolution of 13.2% (FWHM/E) at 59.5 keV. We evaluated the activities of $^{226}$Ra and $^{228}$Th inside the crystal to be $5.9 \pm 0.6$ μBq/kg and $1.6 \pm 0.3$ μBq/kg, respectively, which would indicate a contamination from $^{238}$U and $^{232}$Th at part-per-trillion level. We measured an activity of $0.51 \pm 0.02$ mBq/kg due to $^{210}$Pb out of equilibrium and a α quenching factor of $0.63 \pm 0.01$ at 5304 keV. We illustrate the analyses techniques developed to reject electronic noise in the lower part of the energy spectrum. A cut-based strategy and a multivariate approach indicated a rate, attributed to the intrinsic radioactivity of the crystal, of ∼1 count/day/kg/keV in the [5–20] keV region.

## 1 Introduction

The SABRE experiment seeks to detect the annual modulation of the dark matter interaction rate in NaI(Tl) crystals with sufficient sensitivity to test the long standing DAMA result in a model independent way. Among the best-motivated candidates to solve the dark matter puzzle [1], Weakly Interacting Massive Particles (WIMPs) [2] are thought to exist as a halo within which our Galaxy resides. Despite the decades-long effort of direct detection experiment searches for interactions of such dark matter candidates with different target materials, the DAMA experiment (short for DAMA/NaI and DAMA/LIBRA) [3] is the only one reporting a positive obser-vation. The DAMA evidence of a clear annual modulation, which satisfies the criteria for a model-independent WIMP-induced signal [4], has reached a strong statistical significance of 12.9 $\sigma$ [5], after 20 years of data taking at the Gran Sasso National Laboratory (LNGS), in Italy. While the null result of experiments based on different target materials and other direct detection techniques [6–13] cannot be reconciled with such modulation in the simplest scenario of WIMP-nucleon elastic scattering, a NaI(Tl) annual modulation search similar to DAMA is still crucial to shed light on the origin of the positive signal [14].

[a] e-mail: giulia.dimperio@roma1.infn.it (corresponding author)

Currently running NaI(Tl) dark matter experiments are COSINE-100 [15] at the YangYang Laboratory in South Korea, and ANAIS-112 [16] at the Canfranc Laboratory in Spain. They both feature a mass of about 100 kg of NaI(Tl) scintillating crystals in an underground setup and have so far neither confirmed nor fully disproved the DAMA observation [17–19]. Given the very low count rate and energy featured in dark matter nuclear recoils, its detection is challenged by background from radioactive decays, where the dominating component arises from impurities in the crystals themselves. One key feature of DAMA that has never been rivalled by competing experiments is the background rate which is as low as $\sim 0.7$ count/day/kg/keV in the energy range where the modulation is observed: [1–6] keV [5]. The amplitude of the observed modulation is at the level of 0.01 count/day/kg/keV. Consequently, background levels about two or three times higher than those of DAMA/LIBRA, such as achieved by the ANAIS and COSINE experiments, negatively affect their sensitivity. Indeed these experiments, even after several years of operation, might not be able to resolve all possible scenarios in interpreting the DAMA signal as a dark matter signature.

The SABRE concept and ambitious goal is to obtain an ultra-low background rate in the energy region of interest, namely of the order of 0.1 count/day/kg/keV, that is several times lower than the DAMA/LIBRA level. This challenging goal is achievable by developing high-purity crystals and operating them inside a liquid scintillator veto for active background rejection. This SABRE veto concept has also been adopted by the COSINE experiment.

The SABRE scientific program is currently in a Proof-of-Principle (PoP) phase at LNGS, to assess the radio-purity of the crystals, as well as the acceptance of the liquid scintillator veto. In the longer term, twin detectors of at least 50 kg of NaI(Tl) will be located in the northern and southern hemispheres, namely at LNGS and at the Stawell Underground Physics Laboratory (SUPL) in Australia, located 240 km north-west of Melbourne, where a section of an active gold mine is being converted into a laboratory. The twin locations will help to identify any possible contribution to the modulation from seasonal or site-related effects. The SABRE concept and the PoP setup are described in more detail in Ref. [20].

In this paper we focus on the initial characterization of a relatively large crystal, named NaI-33, that marks a breakthrough after a technical effort of several years. In Sect. 2 we recall the SABRE production strategy for high purity crystals; in Sect. 2.1 we give more details about the production of the crystal NaI-33; in Sect. 3 we describe the underground facility used for the measurements and we detail the data taking campaign; in Sect. 4 we show the main results on the crystal parameters and backgrounds and we also outline the analysis strategy we developed.

## 2 SABRE high purity NaI(Tl) crystals

For many years, the quest for high-purity NaI(Tl) crystals has focused on removing common impurities such as $^{238}$U, $^{232}$Th, $^{40}$K and $^{87}$Rb. One of the $^{40}$K decay channels includes an Auger electron or X-ray cascade with a total energy of $\sim 3.2$ keV, which happens to be in the energy region where the DAMA modulation is observed. Consequently, a large effort was devoted to reduce, or being able to identify, the $^{40}$K background. Eventually, the potassium content in the NaI(Tl) powder and crystals was progressively lowered to the point where the background is now dominated by $^{210}$Pb and cosmogenic-activated isotopes [21], in particular $^{3}$H.

To keep the intrinsic contaminants at a very low level, SABRE developed a method to obtain ultra-pure NaI powder and a clean procedure to grow crystals. More details on this method can be found in Refs. [20,22]. Princeton University and industrial partner Sigma-Aldrich produced ultra-high purity NaI powder, so-called Astro Grade powder, with $^{39}$K levels consistently lower than 10 ppb. The NaI(Tl) crystals were then grown by Radiation Monitoring Devices, Inc. (RMD), using the vertical Bridgman technique [23]. In this method, the powder is placed inside a sealed ampule, reducing the possibility of contamination during the growth phase. In 2015, we successfully grew at RMD a 2 kg crystal with an average $^{39}$K level of $9 \pm 1$ ppb and with $^{87}$Rb upper limit of 0.1 ppb measured by Inductively Coupled Plasma Mass Spectrometry (ICP-MS) [24,25]. Based on these results, a simulation study of the background contributions expected for the SABRE PoP phase can be found in Ref. [26]. The study demonstrates the importance of reducing the background contributions from radioactive decays inside the crystal. The same conclusion can be drawn from the background model presented by COSINE and ANAIS experiments in [27,28], verified against experimental data. However ICP-MS only permits testing for a subset of the backgrounds, namely primordial parents such as U, Th, K and Rb. Cosmogenic isotopes produced after the crystal growth, $^{210}$Pb and other non-primordial parents, if present, can only be mea-sured by direct counting of the crystal as a scintillator in an underground setup. Consequently, the focus in the past years has shifted to the growth of a NaI(Tl) crystal with a sufficient mass to perform a physics run. We have started to acquire signals with a SABRE grade NaI(Tl) crystal for a preliminary background characterization reported in this paper. In the next few months we aim at a more comprehensive characterization with the PoP setup.

### 2.1 SABRE crystal NaI-33

At the end of October 2018, a new NaI(Tl) crystal, named NaI-33, was grown. The powder and the crucible were prepared by the Princeton group, and the crystal was grown in an oven at RMD. The crystal was then cut into an octagon and polished, obtaining a final mass of 3.4 kg. In order to determine the concentration of $^{39}$K, and consequently of $^{40}$K, in the final crystal, samples from the tip and the tail of the ingot were analyzed using ICP-MS at Seastar Chemicals. The average $^{39}$K concentration within the crystal volume was extrapolated to be $4.3 \pm 0.2$ ppb, several times lower than 20 ppb, the upper limit reported for the DAMA/LIBRA crystals [29]. Further details on the powder preparation, the crystal growth and the results on internal radioactivity can be found in Ref. [22].

During May 2019, the crystal NaI-33 was assembled in the SABRE detector module, developed for the PoP phase and described in Ref. [20]. Two Photomultiplier Tubes (PMTs) were directly coupled to each end of the crystal, while its lateral surface was wrapped with reflector, for higher light collection acceptance. For NaI-33, we used Polytetrafluoroethylene (PTFE) as a reflector and 3-inch Hamamatsu R11065-20 PMTs, which were specifically developed to have a low intrinsic radioactivity.[1] We purchased selected units with a quantum efficiency as high as $\sim 35\%$ at 420 nm. The detector assembly was placed inside an air- and light-tight copper enclosure. The cylinder, end-caps, and support rods are made out of low-radioactivity copper. The holders of the crystal and of the PMTs are made out of high-purity Delrin. Being that NaI is highly hygroscopic, all parts were thermally treated prior to assembly, to remove any residual moisture. For the same reason, the assembly procedure was performed entirely under dry nitrogen atmosphere inside a glove box. After sealing, the detector module was flushed with high purity 5.5 N$_2$ gas to avoid humidity and radon. The detector module before the sealing of the copper enclosure can be seen in Fig. 1. In order to limit cosmogenic activation, the crystal was then shipped by surface transportation and delivered on 6 Aug 2019 at LNGS, where it was immediately stored underground.

---

[1] Reference [30] reports the radioactive assessment of the constructively identical model R11410.

## 3 Experimental setup at LNGS

The experimental SABRE setup at LNGS consists of two facilities: a shielded counting system in Hall B and the PoP facility in Hall C. Upon arrival of the NaI-33 crystal, the PoP was not yet ready for operations: consequently, for the first crystal characterization reported in this paper, we used our facility in Hall B. The facility consists of a passive shielding to reduce the $\gamma$ environmental background, made of low radioactivity copper and lead. The shielding is enclosed in a Lexan box that can be sealed and flushed with high-purity nitrogen. In August 2019, we commenced data taking with detector NaI-33 and continued for approximately a year. This time was divided into two periods denoted as Run 1 and Run 2, interleaved by the refurbishment of the shielding performed at the end of February 2020. Shielding during Run 1 featured 5 cm of copper and a layer of lead at least 17.5 cm thick on all sides. The experimental volume inside the shielding allowed for the insertion of two detector modules, which was useful to acquire data with a radioactive source (see Sect. 4.5). During Run 2, the thickness of low radioactivity copper was increased to 10 cm, effectively reducing the background due to the surrounding lead layer. However, this was done at the expense of the experimental volume, that then could only host one detector module. Figure 2 shows the two setups.

We define the presence of a PMT pulse by discriminating the signal over an amplitude threshold corresponding to approximately 1/3 of photoelectron. Data acquisition is triggered by the logical AND of the two PMT pulses occurring within a window of 125 ns. For each event, we acquire the waveforms of each PMT, together with their amplified ($\times 10$) copies. Amplified signals are used for the low energy analysis and saturate above 500 keV energy. The PMT signals without amplification instead provide sensitivity to $\alpha$ decays. Waveforms are digitized at 12 bit and 250 MS/s by CAEN v1720 modules. The acquisition window was set at 5 µs, including a 1.5 µs pre-trigger for baseline evaluation. The trigger rate was $\sim 0.48\,\text{s}^{-1}$.

An offline reconstruction program processes each event and computes physical quantities. PMT pulses are isolated within the acquisition window; the baseline is computed on the pre-trigger samples and subtracted; then the pulse integral is computed. The same operations are performed for each channel, amplified and not, and for the software sum of the two PMT pulses. The event energy estimator is the integral of the sum channel. Pulse Shape parameters are also computed during this phase and are useful to separate scintillation events in the crystal from PMT noise, as well as to distinguish $\alpha$ from $\beta/\gamma$ interactions. Pulse Shape parameters are discussed in detail in Sect. 4.4.

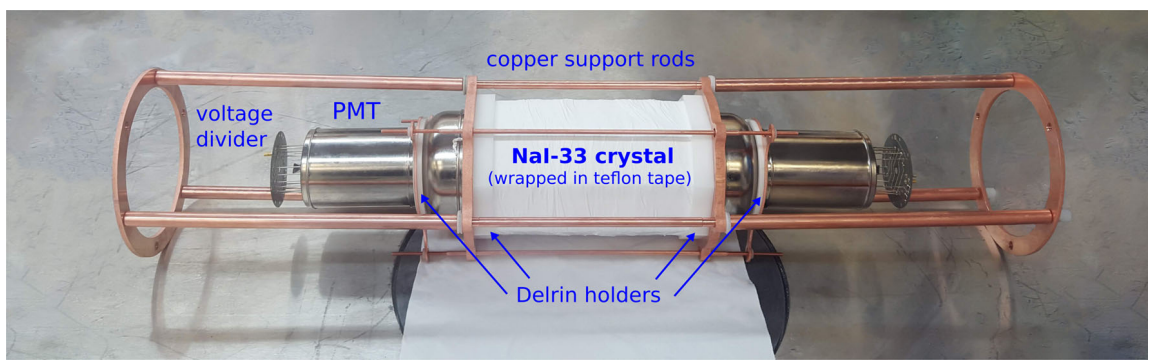

**Fig. 1** NaI-33 assembly in the glovebox, ready for the insertion into the copper enclosure

**Fig. 2** Open passive shielding in the SABRE Hall B area. Left: Run 1 shielding. Right: Run 2 shielding with increased copper layer and detector module in position. The the anti-radon box is not shown here

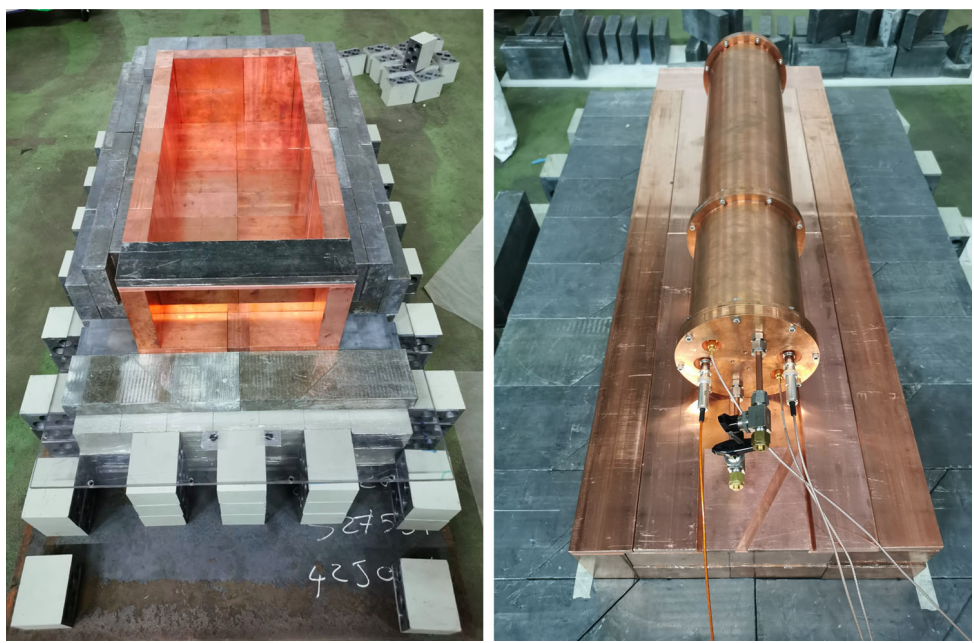

## 4 Data analysis and results

Data acquired during Run 1 were used to measure light yield and energy resolution (Sect. 4.1), and to determine the $\alpha$ rate (Sect. 4.2); however, for the analysis of its evolution over time, all datasets were used. Run 1 included a campaign with a $^{108m}$Ag source (see Sect. 4.5). Run 2 data were used for the background study. In order to build the low energy spectrum, we used an additional dataset, denoted as Run 2-lowRn. These data were acquired while flushing the setup with high-purity nitrogen and, consequently, with a reduced background contribution from Rn and its daughters. A small fraction of the datasets was not included in the analysis due to noisy experimental conditions related to activity in the laboratory. The live time for each run is reported in Table 1.

The characterization of the NaI-33 will proceed within the PoP setup in Hall C, which has recently been commissioned, and it will be the subject of a separate publication.

**Table 1** Runs setup details including live time

| Run         | Live time [d] | Cu thickness [cm] | N$_2$ flushing |
|-------------|---------------|-------------------|----------------|
| Run 1       | 113           | 5                 | Off            |
| Run 2       | 36            | 10                | Off            |
| Run 2-lowRn | 58            | 10                | On             |

### 4.1 Light yield and energy resolution

To characterize the crystal performance, we used a $^{241}$Am calibration source positioned adjacent to the copper enclosure and longitudinally aligned with the crystal centre. We repeated these source runs a few times throughout the data taking. The spectrum acquired with the source is shown in Fig. 3, including a gaussian fit to the 59.5 keV line of $^{241}$Am, on the spectrum built from the sum of the two PMT channels. The crystal light yield and the energy resolution are defined as

$$\text{LY} = \frac{\mu[\text{photoelectrons}]}{E_{peak}[\text{keV}]} \quad (1)$$

$$\frac{\text{FWHM}}{E} = \frac{2.355\,\sigma}{\mu} \quad (2)$$

where $\mu$ and $\sigma$ are the mean and standard deviation from the gaussian fit.

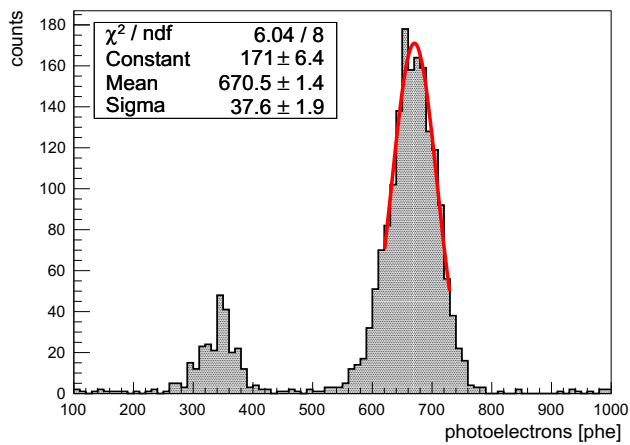

**Fig. 3** NaI-33 calibration spectrum with $^{241}$Am source. The gaussian fit to the 59.5 keV peak is shown in red

The low tail of the $^{241}$Am peak is longer than its high tail due to a small internal crystal contamination of $^{210}$Pb. In fact $^{210}$Pb decays with 80.2% probability emitting a beta with maximum energy 16.9 keV and gamma/electron emissions of 46.5 keV, producing an excess around 50 keV. In order to avoid this bias, we evaluate the energy resolution fitting in an asymmetric interval around the peak of [620, 730] photoelectrons (phe). The resulting energy resolution at this energy is 13.2%. In order to compute the light yield we independently acquired the S.E.R. (Single Electron Response) spectra of each PMT with a 12-bit, 2.5 GS/s oscilloscope. The ratio of the fitted $^{241}$Am peak position to the S.E.R. returns a light yield of $11.1 \pm 0.2$ phe/keV. The error takes into account the systematic uncertainty due to the fact that among calibration runs we could not possibly place the source at exactly the same position. The contribution from the fit to the S.E.R. spectrum was found to be negligible in comparison. For reference, the light yield obtained by other NaI(Tl) experiments ranges as 6–10 phe/keV (DAMA-PhaseII [31]), 14–15.5 phe/keV (COSINE [32]), 12.7–16 phe/keV (ANAIS [33]).

### 4.2 Measurement of the $\alpha$ rate

The identification and counting of $\alpha$ particles has two main aims: (1) a measurement of the internal radioactivity of the crystal in terms of U/Th contamination and $^{210}$Pb contamination out of secular equilibrium; (2) a measurement of the $\alpha$ quenching factor in NaI.

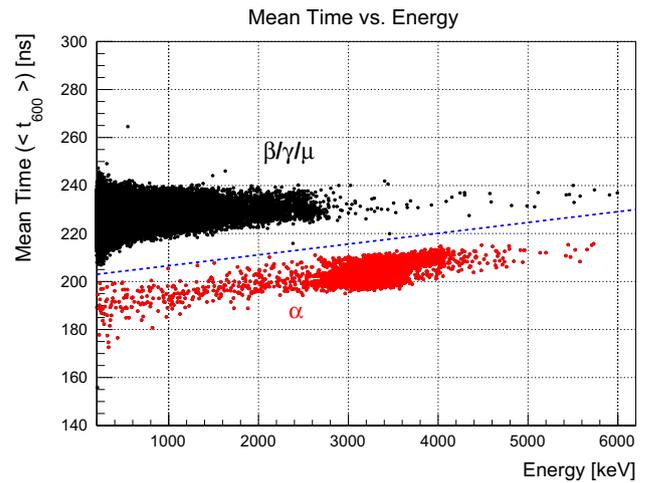

**Fig. 4** Distribution of mean time vs. energy of $\alpha$ (red) and $\beta/\gamma$ (black) events. The blue dotted line shows our typical separation cut

Identification of $\alpha$ particles exploits the Pulse Shape discrimination capability of NaI(Tl) crystals, where signals induced by $\alpha$ decays are faster than $\beta/\gamma$ ones. In particular, the amplitude-weighted mean time $\langle t \rangle_{600}$ (defined in Sect. 4.4), gives excellent separation between $\alpha$ particles and $\beta/\gamma$ particles. This can be seen in Fig. 4 where above 200 keV energy our $\alpha$ selection cut is also shown and gives 100% acceptance. To this end, the calibration of the $\beta$ spectrum was performed using a $^{228}$Th $\gamma$ source. Specifically we exploited the 239 keV line of $^{212}$Pb, the 583 keV and 2615 keV lines of $^{208}$Tl.

In order to evaluate the content of primordial radioactive chains in the crystal, we searched for fast $\beta$-$\alpha$ sequences, called Bismuth-Polonium (Bi-Po). These are present in $^{232}$Th and $^{238}$U chains with atomic mass numbers of 212 and 214, respectively, each with a characteristic decay time. The Bi-Po-212 sequences can be identified in the same digitization windows, because of the very short lifetime of $^{212}$Po (431 ns). Since the lifetime of $^{214}$Po is 237 µs, correlated Bi-Po-214 decays are instead found in event pairs occurring in close succession. Considering the rate of singles is $0.02\,\text{s}^{-1}$ we computed the probability of accidental coincidences to be $\sim 10^{-7}$ ($\sim 10^{-5}$) for the Bi-Po-212 (Bi-Po-214) selection.

We corrected the rates of Bi-Po-214 and Bi-Po-212 sequences for the acceptance of the time-distance cut, respectively 93% and 81%, and we estimated the activity of $^{226}$Ra and $^{228}$Th inside the crystal. These are $5.9 \pm 0.6\,\mu\text{Bq/kg}$ and $1.6 \pm 0.3\,\mu\text{Bq/kg}$, respectively. These uncertainties are statistical only, and are dominant due to the very low number of coincident counts. If the assumption of secular equilibrium for $^{238}$U and $^{232}$Th held, such values would correspond to a part-per-trillion (ppt) level contamination, which is very close to our target [26].

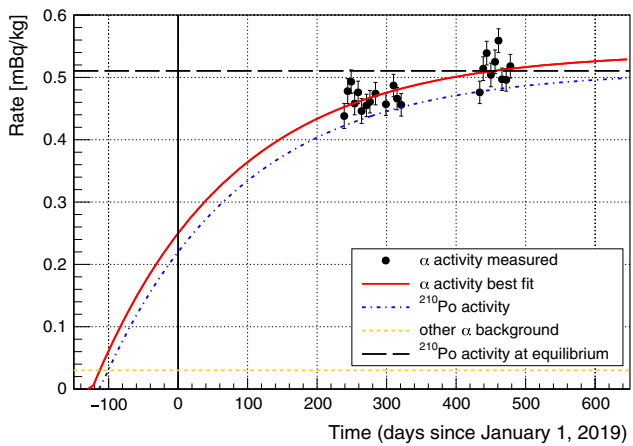

**Fig. 5** $\alpha$ activity in $^{210}$Po energy region as a function of time with 5-days binning. The best fit function to the data (red solid line) is the sum of the build-up function with the mean life of $^{210}$Po (blue dotted line) and a constant component independently determined (orange dotted line). The black dashed line shows the activity of $^{210}$Po at equilibrium

With very few counts ascribed to long-lived radioactive chains, the $\alpha$ spectrum is dominated by the decay of $^{210}$Po. Generally, this originates from an out-of-equilibrium contamination of $^{210}$Pb, accidentally introduced sometime during the manufacture of the crystal (e.g. due to an exposure to radon). We applied a loose energy selection around the $^{210}$Po peak in the $\alpha$ spectrum and measured its activity over a period of a several months with results shown in Fig. 5. We observe an increase well described by a build-up function with the half-life of $^{210}$Po (138.4 days). The fit also includes a constant contribution due to other $\alpha$ decays that satisfy the selection requirement on the energy. This contribution is computed from the result of the two Bi-Po analyses to be 0.03 mBq/kg, including also $^{210}$Po in equilibrium with $^{226}$Ra, and fixed in the fit. The date of the initial $^{210}$Pb contamination is freely determined by the fit to be 9 September 2018 with an uncertainty of 10 days. This date corresponds to the last days before the beginning of the growth process. The activity of out-of-equilibrium $^{210}$Po, and hence of $^{210}$Pb, reaches a plateau of 0.51±0.02 mBq/kg (statistical uncertainty). While such activity is still higher than our target, it is amongst the lowest ever reported after DAMA/LIBRA [27,28].

Measuring the ratio of the $^{210}$Po peak position in the $\beta$-calibrated spectrum over the nominal $\alpha$ energy of 5304 keV, we can estimate the quenching factor for $\alpha$ particles as 0.63± 0.01.

### 4.3 Study of cosmogenic activation

The crystal NaI-33 arrived at LNGS on August 6th and it was exposed to cosmic rays (at sea level) for a total of 279 days. To identify the contribution from cosmogenic activated isotopes, we compared experimental spectra acquired at dif-ferent time periods, namely 29 days, 78 days and 104 days after the arrival of the detector underground. The comparison is done in the energy regions (20–45) keV and (55–80) keV. These are the only intervals where a contribution from cosmogenic isotopes is visible in our experimental spectra. The energy region between 45 and 55 keV was not considered as it is dominated by $^{210}$Pb. The plots in Fig. 6 clearly show a decay of the count rate in the energy regions considered. We attribute this to the decay of the cosmogenic isotopes $^{125}$I (half-life: 59.4 days, main emission: 67.3 keV) and $^{126}$I (half-life: 12.9 days, main emission: 31.8 keV).

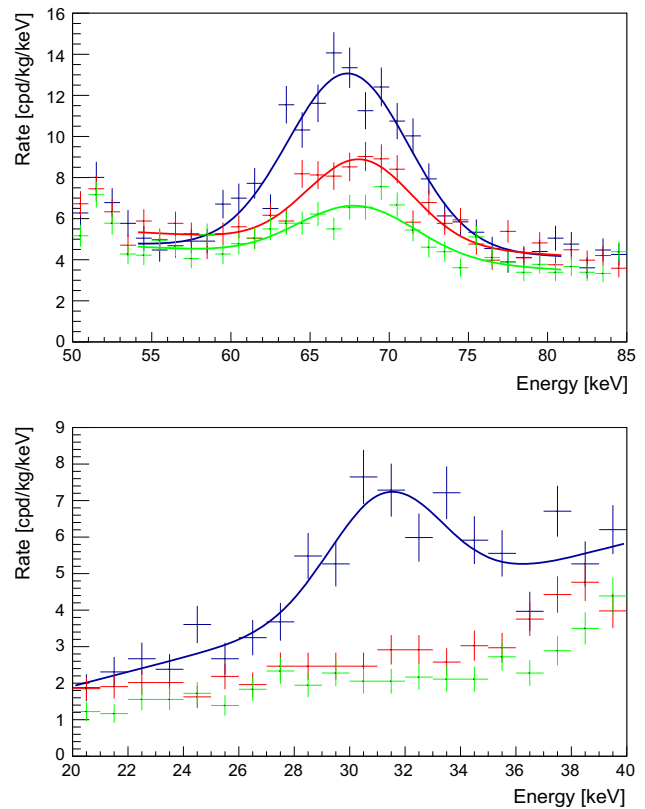

**Fig. 6** Fit to the experimental spectra acquired at different time periods: 29 days (blue), 78 days (red) and 104 days (green) after the arrival of the detector underground. The peak at 67.3 keV (top) is attributed to the decay of $^{125}$I, while the peak at 31.8 keV (bottom) is attributed to the decay of $^{126}$I and is visible only in the measurement at 29 days, given the short decay time

To verify this and quantify the activity of such isotopes, we fitted the two peaks in the three spectra (whenever the peak was visible) with a gaussian plus linear background function and extracted the number of counts in the peaks. This was then compared to the corresponding counts in the Monte Carlo simulated spectra of the isotopes under study. The simulated spectra take into account the decay of both $^{125}$I and $^{126}$I.

Concerning $^{125}$I, the ratios of experimental to simulated data are well in agreement for the spectra at 78 days and 104

**Table 2** Upper limit on the activity of Tellurium cosmogenic isotopes in the NaI-33 crystal, at the time of the crystal arrival underground. For each isotope we indicate the energy of the gamma emission whose $\pm 3\sigma$ interval in the spectrum was used to extract the activity limit

| Isotope | $\gamma$ line [keV] | Limit [mBq/kg] |
| --- | --- | --- |
| $^{127m}$Te | 88.3 | 0.14 |
| $^{125m}$Te | 144.8 | 0.10 |
| $^{123m}$Te | 247.6 | 0.09 |
| $^{121m}$Te | 294.0 | 0.15 |

days, but less so with the spectrum at 29 days. Indeed the value of the ratio is not constant as expected and this might indicate that another isotope with a different decay time is contributing to the peak at 67.3 keV. Considering the spread of the ratios as a measure of the uncertainty with which we can assess the presence of the isotope, this study indicates a $^{125}$I activity at the time of the crystal arrival underground between 1.0 and 1.7 mBq/kg.

A similar estimation was done for $^{126}$I. For this isotope we can only rely on the measurement at 29 days, given the short decay time. The number of counts in the 31.8 keV peak in such spectrum was compared to the corresponding counts in the Monte Carlo spectrum and the resulting activity at the time of the crystal arrival underground is $(3.4 \pm 0.5)$ mBq/kg.

Cosmogenic activation in NaI can also produce isotopes of Tellurium, such as $^{121m}$Te, $^{123m}$Te, $^{125m}$Te, and $^{127m}$Te. We could not directly identify the presence of those isotopes, as their characteristic gamma lines are either not visible on our experimental spectrum or hidden by the above-seen contributions from Iodine cosmogenic activated isotopes. However, by summing the experimental counts in a $\pm 3\sigma$ interval around the energy of an expected gamma line, and assuming the square root of the number of counts as the $1\sigma$ limit on the observable events, we can then compare to the Monte Carlo expectations and extract an upper limit on the activity of a given isotope, at the time of the crystal arrival underground. Table 2 reports the limits we calculated with this method on the experimental spectrum at 29 days.

Converting the cosmogenic activities we derived for Iodine and Tellurium isotopes into production rates at sea level, we obtain values that are in rough agreement with those measured by the ANAIS [21] and COSINE [34] experiments.

### 4.4 Pulse shape parameters

As reported by all NaI(Tl) experiments, the low energy spectrum is populated by events that do not show the correct pulse shape of scintillation light in the crystal, and must be efficiently discarded as noise. We developed a twofold selection strategy for this purpose, detailed in the next sections. More specifically, we report a cut-based method (Sect. 4.5) and a more sophisticated Boosted Decision Tree approach (Sect. 4.6). The results of these analyses are given in Sect. 4.7.

We define here the pulse shape variables common to both approaches; some of them are the same or similar to the ones also exploited by other NaI experiments [31,33,35]. In the definitions of the variables, $h_i$ and $t_i$ are the pulse amplitude in mV and the time of the $i$−th sample (starting from the trigger, $t_0$), respectively. $h_{max}$ is the absolute value of the maximum pulse amplitude in mV. $C_{(t_i,t_f)}$ is defined as the pulse area between $t_i$ and $t_f$ in nanoseconds. $E_0$ and $E_1$ correspond to the energy measured using the two PMTs in the 1000 ns following the trigger.

$$\text{Amplitude weighted mean time } \langle t \rangle_{600} = \frac{\sum_{t_i < 600 \text{ ns}} h_i t_i}{\sum_{t_i < 600 \text{ ns}} h_i} \quad (3)$$

$$\text{Charge over maximum CoM} = \frac{C_{(0,1000)}}{h_{max}} \quad (4)$$

$$\text{Tail-to-total pulse shape } X_1 = \frac{C_{(100,600)}}{C_{(0,600)}} \quad (5)$$

$$\text{Head-to-total pulse shape } X_2 = \frac{C_{(0,50)}}{C_{(0,600)}} \quad (6)$$

$$\text{Asymmetry } A = \frac{E_0 - E_1}{E_0 + E_1} \quad (7)$$

$$\text{Head-to-middle pulse shape } C_2/C_1 = \frac{C_{(200,400)}}{C_{(0,200)}} \quad (8)$$

$$\text{Middle-to-tail pulse shape } C_3/C_2 = \frac{C_{(400,600)}}{C_{(200,400)}} \quad (9)$$

$$\text{Time Variance Var}[t] = \langle t^2 \rangle_{600} - \langle t \rangle_{600}^2 \quad (10)$$

$$\text{Skewness Skw}[t] = \frac{\langle t^3 \rangle_{2000}}{(\langle t^2 \rangle_{2000})^{3/2}} \quad (11)$$

$$\text{Kurtosis Krt}[t] = \frac{\langle t^4 \rangle_{2000}}{(\langle t^2 \rangle_{2000})^2} - 3 \quad (12)$$

$$\text{Combination of } X_1 \text{ and } X_2 \text{ ES} = \frac{1 - (X_2 - X_1)}{2} \quad (13)$$

$$\text{Combination of Skw}[t] \text{ and Krt}[t] \text{ SK} = (a \cdot \text{Skw}[t] - b \cdot \text{Krt}[t] - c)^2 + (\text{Krt}[t] - d)^2 \quad (14)$$

where $a = 30$, $b = 5.2$, $c = 50$, and $d = 2$, as fixed from the analysis of the $^{108m}$Ag run data. In addition to the variables defined above, we define a Kolmogorov–Smirnov distance from an $^{241}$Am reference pulse, and the number of clusters (NC). A cluster is defined by the waveform exceeding an amplitude threshold corresponding to ∼1 phe. Accepting only events with NC > 2 in each of the PMTs, we eliminate accidental coincidences of single photo-electrons in opposing PMTs and events originated in the PMTs due to their intrinsic radioactivity.

### 4.5 Cut-based analysis

We define here the selection criteria adopted for low energy noise rejection, tuned on data acquired with a $^{108m}$Ag source.

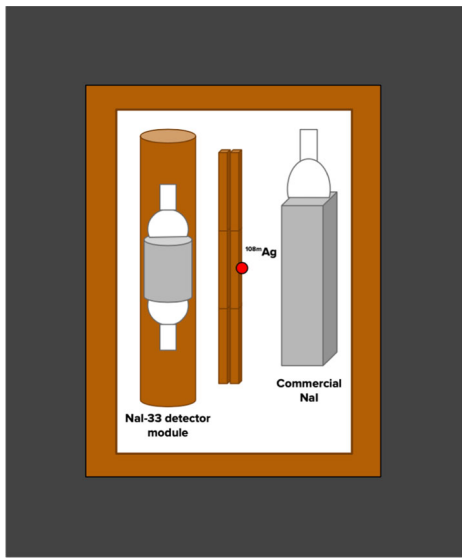

**Fig. 7** Schematic view of the experimental setup used for the $^{108m}$Ag source campaign. It is shown the passive shielding of lead and copper, with the NaI-33 detector module and the commercial NaI detector. $\gamma$ rays emitted from the $^{108m}$Ag source, identified by a red circle, are degraded via a copper layer

The $^{108m}$Ag source produces three $\gamma$ rays of 434 keV, 614 keV, and 723 keV. We exploit these simultaneous emissions by placing a commercial NaI detector along NaI-33 inside the shielding (see Fig. 7). Requiring coincidence between the two detectors within 125 ns and using a copper layer as energy degrader between the source and the NaI-33 crystal, we obtain a continuum of low energy scintillation events. We have evaluated the contamination of the sample by accidental coincidences and found it to be negligible.

We apply cuts on the variables defined in Eqs. 3–10, 13–14, on the number of clusters and on the Kolmogorov-Smirnov distance. The selection criterion on each variable is chosen to conserve at least 95% of signal events with energy within (1–20) keV (with the exception of Eq. 14, where we preserve only 68% of events). We show in Fig. 8 the distributions of a few of the shape parameters in the source data. The values corresponding to the selection criteria are shown as red lines.

The combined acceptance of all the cuts as a function of energy and the resulting low-energy spectrum will be presented below (Sect. 4.7) in comparison with those obtained with the boosted decision tree analysis which will now be described.

### 4.6 Boosted decision tree analysis

In order to improve the separation between noise and scintillation events, we implemented a multivariate analysis using Boosted Decision Trees (BDT) [36,37], with a focus on low energy events up to 100 keV.

The BDT algorithm has to be trained using a sample of true scintillation events as "signal" and a sample of noise event as "background". For the signal training we used the $^{108m}$Ag data, selecting the events in coincidence with the commercial crystal, weighted in order to match the energy distribution expected for the crystal background.

For the background training we used NaI-33 data in the 1–10 keV energy region, which is dominated by noise. We evaluate from the cut-based analysis that the contamination of signal events in the background training set is about 5% and consider this bias reasonably small. As cross-check, we also trained the BDT with a background set obtained by inverting the selection described in Sect. 4.5, achieving comparable results.

The analysis was performed by training two separate BDTs, which we refer to as $BDT_1$ and $BDT_2$. The input variables for $BDT_1$ are those defined in Eqs. 3–10, while for $BDT_2$ we add also Skewness and Kurtosis (11 and 12), similarly to what is done in [15]. As for the cut-based analysis, we evaluate the selection on the BDT variables using the data with the $^{108m}$Ag source.

The BDT classifier output is defined in the range $[-1, 1]$ where lower scores are assigned to noise-like events and higher scores to signal-like events. Figure 9 shows the output as a function of the energy for the source data.

We define an energy-dependent threshold for the two BDTs defined as following:

$$\begin{cases} BDT_{1(2)} > \frac{a_{1(2)}}{\log_{10}(20)} \log_{10}(E) - b_{1(2)} & : \text{if } (1 \leq E < 20) \text{ keV} \\ BDT_{1(2)} > c_{1(2)} & : \text{if } E \geq 20 \text{ keV} \end{cases}$$
(15)

where $a_i$, $b_i$ and $c_i$ are chosen in order to ensure an acceptance for each BDT of $\sim$50% (99.8%) in the energy range [1–3] keV ([20–100] keV).

We then combine the two BDT algorithms by selecting as scintillation events only those passing both thresholds, with the results described in the next section.

### 4.7 Comparison of analysis procedures

The results of the cut-based analysis and the BDT analysis are shown in Figs. 10 and 11. Figure 10 compares the acceptance as a function of energy below 60 keV. The acceptance is measured using the $^{108m}$Ag data, by computing the fraction of events passing the selection. Figure 11 shows the low energy spectrum before and after noise rejection. This spectrum corresponds to a live time of 58 days. The spectra obtained with the two methods are corrected for their respective acceptance. After noise rejection the average count rate in the 5–20 keV energy range is close to 1 count/day/kg/keV.

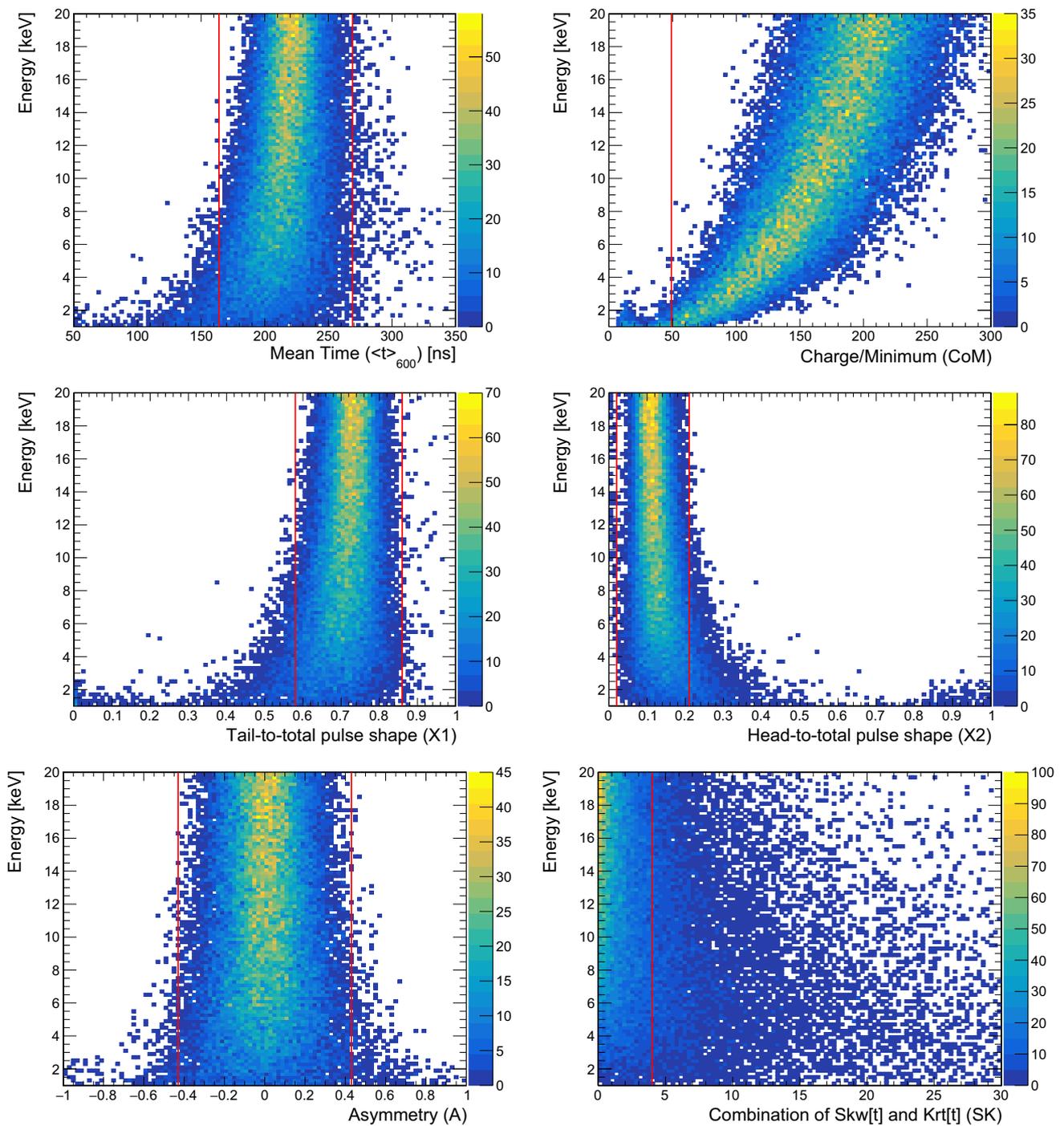

**Fig. 8** Distributions of some of the shape parameters in the data with a $^{108m}$Ag source. Red lines indicate the thresholds of the selection criteria

The count rate below 5 keV will be evaluated operating the crystal in the PoP setup with more favourable noise conditions.

The multivariate analysis shows a promising performance: the acceptance is significantly higher with respect to the cut-based approach. At the same time, the energy spectra of events selected with the two methods, after being corrected for the acceptance, compare well above 8 keV whereas at lower energies the BDT approach grants a slightly higher noise rejection.

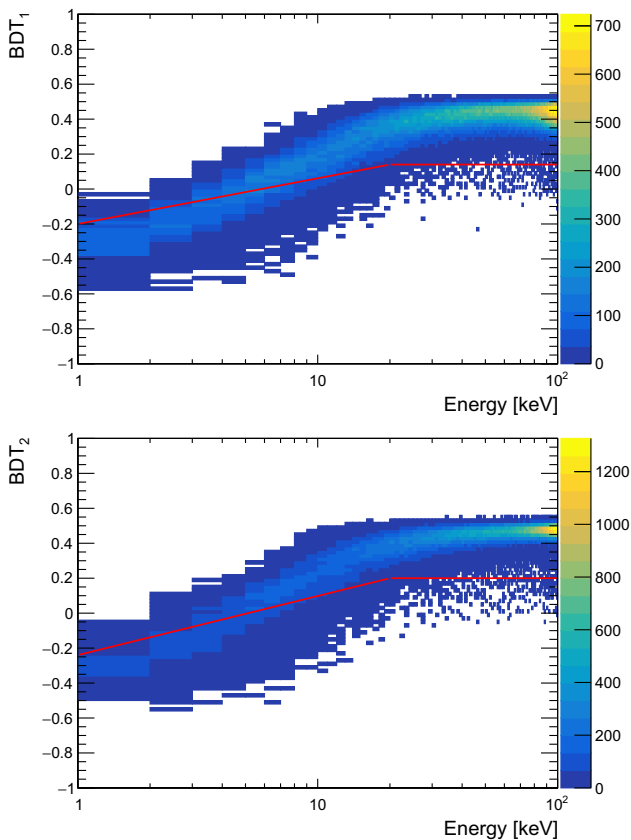

**Fig. 9** $BDT_1$ (top) and $BDT_2$ (bottom) classifiers output for the $^{108m}$Ag data, with superimposed energy dependent threshold in red

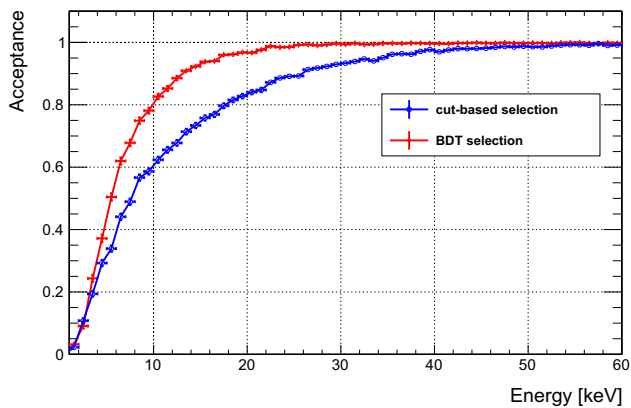

**Fig. 10** Acceptance vs energy for the cut-based analysis and the multivariate BDT analysis

## 5 Conclusions

The NaI-33 crystal is the most radiopure crystal produced so far by the SABRE collaboration, with a size suitable for the PoP phase and close to the target of 5 kg for the full-scale experiment. The $^{39}$K content measured with ICP-MS is $4.3 \pm 0.2$ ppb, significantly lower than all values reported by experiments DAMA, which quotes a 20 ppb upper limit

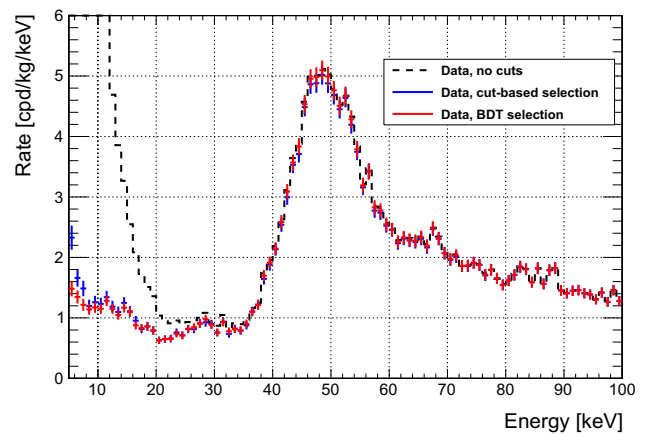

**Fig. 11** NaI-33 low energy spectrum before (black dashed line) and after noise rejection. We show the result of the cut-based (blue points) and multivariate BDT (red points) analyses. The blue and red histograms are corrected by the respective cut acceptance

[29], COSINE-100 and ANAIS-112, which have both 17 ppb at best [27,28]. We reported here the results from a first characterization performed using the SABRE passive shielding setup in Hall B of LNGS.

The crystal scintillation properties were measured with the 59.5 keV line of an $^{241}$Am source. Our detector shows a competitive light yield, as high as $11.1 \pm 0.2$ phe/keV, and an energy resolution of 13.2% (FWHM/E).

The presence of trace elements from primordial radioactive chains was evaluated using Bi-Po delayed coincidence analysis and found to be at the part-per-trillion (ppt) level. However, $^{210}$Pb out of secular equilibrium is observed, and is one of the major sources of background at low energy for NaI(Tl) experiments. We studied the build-up of $^{210}$Po over time which allowed us to estimate the present activity of $^{210}$Pb to be $0.51 \pm 0.02$ mBq/kg. We also measured an $\alpha$ quenching factor of $0.63 \pm 0.01$ at the $^{210}$Po decay energy $E_\alpha = 5304$ keV.

We estimated the activity of Iodine cosmogenic isotopes $^{125}$I (1.0–1.7 mBq/kg) and $^{126}$I ($3.4 \pm 0.5$ mBq/kg) at the time of the arrival of crystal NaI-33 underground, by studying the decay of the count-rate in the corresponding gamma lines and comparing with Monte Carlo simulations. Only upper limits on the activity of Tellurium cosmogenic isotopes could be derived, as no characteristic gamma emissions from these isotopes were visible in our spectra.

We employed two pulse-shape-driven analysis techniques to reject noise at low energy: a cut-based analysis and a multivariate Boosted Decision Tree. The latter shows a significantly higher acceptance. The acceptance-corrected energy spectra obtained with the two methods are found to be compatible above ∼8 keV. Below this energy, the spectra are similar though the multivariate analysis shows a slightly better noise rejection. Part of the rate can be attributed to residual noise events, leaving room for improvements in the rejection in the region around and below 5 keV.

We will continue the characterization of crystal NaI-33 in the PoP setup, equipped with the active veto system. The goal in this phase will be to perform direct counting of decays from $^{40}$K and possibly $^{22}$Na, and to build a complete background model. The commissioning was successfully completed in July 2020 and the measurement of crystal NaI-33 inside the liquid scintillator started in August.

The results already achieved in the passive shielding and presented in this work show that the average interaction rate in the 5–20 keV energy range is close to 1 count/day/kg/keV. This, and other results presented in this paper, show a breakthrough achievement for the SABRE project. The PoP setup is expected to further improve the current understanding of the NaI-33 background toward the SABRE full scale experiment design.

**Acknowledgements** The SABRE program is supported by funding from INFN (Italy), NSF (USA), and ARC (Australia Grants: LE170100162, LE16010080, DP170101675, LP150100075, CE200100008). We thank the Director and staff of the Laboratori Nazionali del Gran Sasso for their support, in particular Dr. Matthias Laubenstein. We thank Sergio Parmeggiano from INFN Milano.

**Data Availability Statement** This manuscript has no associated data or the data will not be deposited. [Authors' comment: The datasets generated during and/or analysed during the current study are available from the corresponding author on reasonable request.]

### References

1. G. Bertone, D. Hooper, Rev. Mod. Phys. **90**, 045002 (2018). https://doi.org/10.1103/RevModPhys.90.045002
2. G. Arcadi et al., Eur. Phys. J. C **78**(3), 203 (2018). https://doi.org/10.1140/epjc/s10052-018-5662-y
3. R. Bernabei et al., Eur. Phys. J. C **73**, 2648 (2013). https://doi.org/10.1140/epjc/s10052-013-2648-7
4. K. Freese, J. Frieman, A. Gould, Phys. Rev. D **37**, 3388 (1988). https://doi.org/10.1103/PhysRevD.37.3388
5. R. Bernabei et al., Nucl. Phys. Atom. Energy **19**(4), 307 (2019). https://doi.org/10.15407/jnpae2018.04.307
6. K. Abe et al., Phys. Rev. D **97**, 102006 (2018). https://doi.org/10.1103/PhysRevD.97.102006
7. M. Kobayashi et al., Phys. Lett. B **795**, 308 (2019). https://doi.org/10.1016/j.physletb.2019.06.022
8. D.S. Akerib et al., Phys. Rev. D **98**, 062005 (2018). https://doi.org/10.1103/PhysRevD.98.062005
9. E. Aprile et al., Phys. Rev. Lett. **118**, 101101 (2017). https://doi.org/10.1103/PhysRevLett.118.101101
10. E. Aprile et al., Phys. Rev. Lett. **121**, 111302 (2018). https://doi.org/10.1103/PhysRevLett.121.111302
11. P. Agnes et al., Phys. Rev. Lett. **121**, 081307 (2018). https://doi.org/10.1103/PhysRevLett.121.081307
12. R. Agnese et al., Phys. Rev. D **97**, 022002 (2018). https://doi.org/10.1103/PhysRevD.97.022002
13. A.H. Abdelhameed et al., Phys. Rev. D **100**, 102002 (2019). https://doi.org/10.1103/PhysRevD.100.102002
14. F. Froborg, A.R. Duffy, J. Phys. G Nucl. Part. Phys. **47**(9), 094002 (2020). https://doi.org/10.1088/1361-6471/ab8e93
15. P. Adhikari et al., Nature **564**(7734), 83 (2018). https://doi.org/10.1038/s41586-018-0739-1
16. I. Coarasa et al., Eur. Phys. J. C **79**(3), 233 (2019). https://doi.org/10.1140/epjc/s10052-019-6733-4
17. G. Adhikari et al., Phys. Rev. Lett. **123**, 031302 (2019). https://doi.org/10.1103/PhysRevLett.123.031302
18. J. Amaré et al., Phys. Rev. Lett. **123**, 031301 (2019). https://doi.org/10.1103/PhysRevLett.123.031301
19. J. Amare et al., (2021). arXiv:2103.01175
20. M. Antonello et al., Eur. Phys. J. C **79**(4), 363 (2019). https://doi.org/10.1140/epjc/s10052-019-6860-y
21. J. Amare et al., Eur. Phys. J. C **79**(5), 412 (2019). https://doi.org/10.1140/epjc/s10052-019-6911-4
22. B. Suerfu et al., Phys. Rev. Res. **2**(1), 013223 (2020). https://doi.org/10.1103/PhysRevResearch.2.013223
23. P.W. Bridgman, Proc. Am. Acad. Arts Sci. **60**(6), 305 (1925)
24. I.J. Arnquist, E.W. Hoppe, Nucl. Instrum. Methods A **851**, 15 (2017). https://doi.org/10.1016/j.nima.2017.01.064
25. Seastar Chemicals, Private communication
26. M. Antonello et al., Astropart. Phys. **106**, 1 (2019). https://doi.org/10.1016/j.astropartphys.2018.10.005
27. P. Adhikari et al., Eur. Phys. J. C **78**(6), 490 (2018). https://doi.org/10.1140/epjc/s10052-018-5970-2
28. J. Amaré et al., Eur. Phys. J. C **76**(8), 429 (2016). https://doi.org/10.1140/epjc/s10052-016-4279-2
29. R. Bernabei et al., Nucl. Instrum. Methods A **592**(3), 297 (2008). https://doi.org/10.1016/j.nima.2008.04.082
30. E. Aprile et al., Eur. Phys. J. C **75**(11), 546 (2015). https://doi.org/10.1140/epjc/s10052-015-3657-5
31. R. Bernabei et al., J. Instrum. **7**(03), P03009 (2012). https://doi.org/10.1088/1748-0221/7/03/p03009
32. G. Adhikari et al., Eur. Phys. J. C **78**(2), 107 (2018). https://doi.org/10.1140/epjc/s10052-018-5590-x
33. J. Amaré et al., Eur. Phys. J. C **79**(3), 228 (2019). https://doi.org/10.1140/epjc/s10052-019-6697-4
34. EB de Souza et al., Astropart. Phys. **115**, 102390 (2020). https://doi.org/10.1016/j.astropartphys.2019.102390
35. G. Adhikari et al., JCAP **06**, 048 (2019). https://doi.org/10.1088/1475-7516/2019/06/048
36. J. Friedman, Ann. Stat. **29**, 1189 (2001). https://doi.org/10.2307/2699986
37. H.J. Yang, B.P. Roe, J. Zhu, Nucl. Instrum. Methods A **555**, 370 (2005). https://doi.org/10.1016/j.nima.2005.09.022